\documentclass[12pt]{article}
\usepackage[utf8]{inputenc}
\usepackage{geometry}
\usepackage{natbib}
\setcitestyle{maxcitenames=2}
\usepackage{listings}
\usepackage{indentfirst}
\usepackage{fancyhdr}
\usepackage[english]{babel}
\usepackage[]{amsthm}
\usepackage[]{amssymb}
\usepackage{amsmath}
\usepackage{graphicx}

\usepackage{epstopdf}
\AppendGraphicsExtensions{.eps}
\usepackage{float}
\usepackage{bm}
\usepackage[shortlabels]{enumitem}
\usepackage{tcolorbox}
\usepackage{hyperref}
\usepackage{tikz}
\usetikzlibrary{matrix}
\makeatletter

\newcommand{\Rmnum}[1]{\expandafter\@slowromancap\romannumeral #1@}
\makeatother

\begin{document}
 
\title{The Implied Views of Bond Traders on the Spot Equity Market}
\author{Yifan He\footnote{Department of Mathematics and Statistics, Texas Tech University, Lubbock, TX 79409-1042, USA,\\email: \href{mailto: yifan.he@ttu.edu}{yifan.he@ttu.edu}}\and Yuan Hu\footnote{Department of Mathematics, University of California San Diego, La Jolla, CA 92903, USA,\\email: \href{mailto:yuh099@ucsd.edu}{yuh099@ucsd.edu}}\and Svetlozar Rachev\footnote{Department of Mathematics and Statistics, Texas Tech University, Lubbock, TX 79409-1042, USA,\\email: \href{mailto: zari.rachev@ttu.edu}{zari.rachev@ttu.edu}}}

\date{October 17, 2023}
\maketitle

\begin{abstract}
This study delves into the temporal dynamics within the equity market through the lens of bond traders. Recognizing that the riskless interest rate fluctuates over time, we leverage the Black-Derman-Toy model to trace its temporal evolution. To gain insights from a bond trader's perspective, we focus on a specific type of bond: the zero-coupon bond. This paper introduces a pricing algorithm for this bond and presents a formula that can be used to ascertain its real value. By crafting an equation that juxtaposes the theoretical value of a zero-coupon bond with its actual value, we can deduce the risk-neutral probability. It is noteworthy that the risk-neutral probability correlates with variables like the instantaneous mean return, instantaneous volatility, and inherent upturn probability in the equity market. Examining these relationships enables us to discern the temporal shifts in these parameters. Our findings suggest that the mean starts at a negative value, eventually plateauing at a consistent level. The volatility, on the other hand, initially has a minimal positive value, peaks swiftly, and then stabilizes. Lastly, the upturn probability is initially significantly high, plunges rapidly, and ultimately reaches equilibrium.
\end{abstract}

\section{Introduction}
Our study begins with an exploration of the market, as characterized by the SPDR S\&P 500 ETF Trust (SPY). We develop a binomial tree for this market, in line with the conventions of previous work on binomial option pricing \citep{hu2022market}, ensuring the preservation of the market's mean and natural upturn probability. Following this foundational understanding, we transition into the evaluation of the term structure of interest rates (TSIR).\par
A common approach within the TSIR involves drawing conclusions based on treasury data and subsequent hypotheses. It is noteworthy that standard TSIR models frequently assign risk-neutral probabilities as follows: $(\tilde{p},\tilde{q}) = (\frac{1}{2},\frac{1}{2})$. However, this assumption is challenged by real-world data from the financial market, as highlighted by \citep{hu2022market} and \citep{shreve2004stochastic}. A possible avenue then emerges: to decipher the risk-neutral dynamics, even if only through a largely theoretical model, that align with bond prices. By leveraging such probabilities, models such as \citep{hu2022market} can offer a valuation of derivatives on the TSIR.\par
In the context of this paper, our meticulous preservation of the equity market's mean, natural upturn probability, and volatility in our bond valuation opens up an avenue to discern bond traders' perspective on these parameters. This perspective is in contrast to the implied probabilities perceived by option traders, a nuance previously overlooked by traditional TSIR models. The crux of this paper, therefore, is to explore the nuances of the equity market through the lens of bond traders. This research is organized into three pivotal sections.\par
First, we elucidate the Black-Derman-Toy (BDT) model, tracing its historical significance in mathematical finance. Our emphasis remains primarily on the discrete incarnation of the BDT model, which offers insights into the market's temporal riskless interest rate transformations. For the computation of the associated BDT model coefficients, we harness the prior 252 days of U.S. 10-year treasury bond interest rate data, which are processed using MATLAB.\par
Next, our study pivots to the vantage point of bond traders, introducing the concept of the zero-coupon bond. Through a binomial model, we endeavor to derive a systematic pricing algorithm for this bond; this is followed by a method that can be used to ascertain its real value. By establishing a relationship between the theoretical and real values of the zero-coupon bond, we identify the risk-neutral probability. Given that a simplistic risk-neutral probability assumption of $(\tilde{p},\tilde{q}) = (\frac{1}{2},\frac{1}{2})$ might yield unrealistic outcomes\footnote{An in-depth discussion on this topic can be found in \citep{hu2022market}.}, we expand this concept to derive parameters such as the instantaneous mean return, volatility, and natural upturn probability. This is further bolstered with simulated data, which are used to interpret these parameters over time.\par
In our conclusion, we discuss insights into the evolutionary trajectory of the equity market over the span of our study.

\section{BDT Model}
\subsection{The History of the BDT Model}
The BDT model, named after its creators, Black, Derman, and Toy, originated within Goldman Sachs during the 1980s. Initially conceived for the firm's internal utilization, its significance garnered broader recognition following its publication in the \textit{Financial Analysts Journal} in 1990 \citep{black1990one}. Derman offers a first-hand account of the model's evolution in his 2004 memoir \cite{derman2004my}.\par
Within the realm of mathematical finance, the BDT model is highly regarded as a short-rate model that is instrumental in pricing bond options, swaptions, and various other interest rate derivatives. It is different from other finance models in that it is a one-factor model, meaning that the short rate serves as the sole stochastic determinant for the future trajectory of all interest rates. The BDT model pioneered the amalgamation of the mean-reverting behavior of the short rate with the log-normal distribution, a paradigm shift underscored by \citep{buetow2001impact}. Its applicability and relevance persist today, as noted by \citep{fabozzi2007fixed}.

\subsection{Theoretical Support}
In \citep{shreve2004stochastic}, one can find a presentation of the BDT model in the discrete case. In the present paper, we will also consider this form of the BDT model. Because the interest rate in the BDT model changes over time, we can consider the interest rate as a risky asset $\mathbb{R}$ and suppose that at every point in time, the interest rate will strictly go up or not. Additionally, we suppose that the interval between two adjacent points in time is $\Delta = 1/252$\footnote{This is based on the standard notion that a typical year comprises 252 business days, making the time interval $1/252$, with the units being years.}. Now, we define the interest rate return\footnote{$R_{n\Delta}$ signifies the value of the interest rate at time $n\Delta$. Similar notation conventions apply elsewhere.} in the period $[n\Delta, (n+1)\Delta]$:
\begin{align}
r_{(n+1)\Delta} = \frac{R_{(n+1)\Delta} - R_{n\Delta}}{R_{n\Delta}}.\label{eq:01}
\end{align}
For every $n$ in the range $0,\cdots, N-1$\footnote{In our study, we utilize data spanning 252 days to estimate each parameter and refine our BDT model. Consequently, $N = 1/\Delta = 252$.}, the interest rate return can be described as follows:
\begin{align}
r_{(n+1)\Delta} = 
\begin{cases}
u\Delta &\textrm{with probability }p,\\
d\Delta &\textrm{with probability }1-p,
\end{cases}\label{eq:02}
\end{align}
where the sequence of returns, denoted by $r_{n\Delta}$\footnote{When $r_{(n+1)\Delta}$ equals $u\Delta$, the interest rate exhibits a clear upward trajectory. However, when $r_{(n+1)\Delta}$ is equal to $d\Delta$, the interest rate does not strictly ascend. By design, $u>d$ should hold true.} for $n=1,\cdots, N$, comprises identically distributed binary random variables.\par
Furthermore, let the expected return of $r_{(n+1)\Delta}$ be represented as $\mathbb{E}[r_{(n+1)\Delta}] = \mu_{(n+1)\Delta}^{\textrm{rate}} = \mu^{\textrm{rate}}\Delta$, where $\mu^{\textrm{rate}}$ stands for the instantaneous mean return of the interest rate. Therefore, the relationship becomes
\begin{align*}
\mu_{(n+1)\Delta}^{\textrm{rate}} = \mu^{\textrm{rate}}\Delta = pu\Delta + (1-p)d\Delta,
\end{align*}
which implies that
\begin{align}
\mu^{\textrm{rate}} = pu + (1-p)d.\label{eq:03}
\end{align}
Now, let the variance of $r_{(n+1)\Delta}$ be represented as follows:
\begin{align*}
\mathbb{V}[r_{(n+1)\Delta}] = (\sigma_{(n+1)\Delta}^\textrm{rate})^2 = (\sigma^{\textrm{rate}})^2\Delta,
\end{align*}
where $(\sigma^{\textrm{rate}})^2$ is the instantaneous variance of the interest rate. By introducing
\begin{align*}
\nu^{\textrm{rate}}_{\Delta} = \frac{\sigma^{\textrm{rate}}}{\sqrt{\Delta}},
\end{align*}
we can obtain
\begin{align}
(\sigma_{(n+1)\Delta}^\textrm{rate})^2 = (\sigma^{\textrm{rate}})^2\Delta = (\nu^{\textrm{rate}}_{\Delta})^2\Delta^2.\label{eq:04}
\end{align}
Given that,
\begin{align}
\mathbb{V}[r_{(n+1)\Delta}] = \mathbb{E}[r_{(n+1)\Delta}]^2 - \mathbb{E}^2[r_{(n+1)\Delta}].\label{eq:05}
\end{align}
By combining (\ref{eq:04}) and (\ref{eq:05}), we can derive
\begin{align}
(\sigma_{(n+1)\Delta}^\textrm{rate})^2 = (\nu^{\textrm{rate}}_{\Delta})^2\Delta^2 &= pu^2\Delta^2 + (1-p)d^2\Delta^2 - [pu\Delta + (1-p)d\Delta]^2\\
&= pu^2 + (1-p)d^2 - p^2 u^2 - 2p(1-p)ud - (1-p)^2d^2\\
&= p(1-p)u^2 + p(1-p)d^2 - 2p(1-p)ud\\
&= p(1-p)(u-d)^2.\label{eq:09}
\end{align}
Since $u > d$, $\nu_{\Delta} = \sqrt{p(1-p)}(u-d)$.\par
By combining the expressions for $\mu^{\textrm{rate}}$ and $\nu_{\Delta}^{\textrm{rate}}$ in (\ref{eq:03}) and (\ref{eq:09}), we obtain the following system of equations:
\begin{align}
\begin{cases}
\mu^{\textrm{rate}} &= pu + (1-p)d,\\
\nu_{\Delta}^{\textrm{rate}} &= \sqrt{p(1-p)}(u-d).
\end{cases}\label{eq:10}
\end{align}
If we treat $\mu^{\textrm{rate}}$ and $\nu_{\Delta}^{\textrm{rate}}$ as constants and $u$ and $d$ as unknown variables, the solution of (\ref{eq:10}) should be
\begin{align}
\begin{cases}
u &= \mu^{\textrm{rate}} + \sqrt{\frac{1-p}{p}}\nu_{\Delta}^{\textrm{rate}},\\
d &= \mu^{\textrm{rate}} - \sqrt{\frac{p}{1-p}}\nu_{\Delta}^{\textrm{rate}}.
\end{cases}\label{eq:11}
\end{align}
Now, by combining (\ref{eq:01}), (\ref{eq:02}), and (\ref{eq:11}), we obtain
\begin{align}
R_{(n+1)\Delta} &= 
\begin{cases}
R_{n\Delta}(1+u\Delta)&\textrm{ with probability }p,\\
R_{n\Delta}(1+d\Delta)&\textrm{ with probability }1-p,
\end{cases}\\
&=
\begin{cases}
R_{n\Delta}\left(1 + \mu^{\textrm{rate}}\Delta + \sqrt{\frac{1-p}{p}}\nu_{\Delta}^{\textrm{rate}}\Delta\right) &\textrm{ with probability }p,\\
R_{n\Delta}\left(1 + \mu^{\textrm{rate}}\Delta - \sqrt{\frac{p}{1-p}}\nu_{\Delta}^{\textrm{rate}}\Delta\right) &\textrm{ with probability }1-p,\label{eq:13}
\end{cases}
\end{align}
for $n=0,\cdots, N-1$.\par
Next, considering the format of the BDT model on page 172 in \citep{shreve2004stochastic},
\begin{align*}
R_{n}(\omega_1\cdots\omega_n) = a_nb_n^{\#H(\omega_1\cdots\omega_n)},
\end{align*}
where
\begin{itemize}
\item $\omega_i$ for $i-1,\cdots, n$ represents the resulting change in the interest rate. $\omega_i$ can be either $H$ or $T$. If $\omega_i = H$, the interest rate increases; if $\omega_i = T$, the interest rate does not increase.
\item $\#H(\omega_1\cdots\omega_n)$ counts the occurrences of $H$ in the given $n$ periods, representing the number of times the interest rate increases.
\item $a_n$ and $b_n$ are coefficients used to calibrate the model.
\end{itemize}
Following the form of $a_n$ and $b_n$ in \citep{shreve2004stochastic}, we assume that $a_n = R_0/c_1^n$ and $b_n = c_2$, where $R_0$ is the interest rate at time 0, while $c_1$ and $c_2$ are constants. By combining this assumption with (\ref{eq:13}), we deduce the following:
\begin{align}
\frac{c_2}{c_1} &= 1+\mu^{\textrm{rate}}\Delta + \sqrt{\frac{1-p}{p}}\nu_{\Delta}^{\textrm{rate}}\Delta, \label{eq:14}\\
\frac{1}{c_1} &= 1+\mu^{\textrm{rate}}\Delta - \sqrt{\frac{p}{1-p}}\nu_{\Delta}^{\textrm{rate}}\Delta.\label{eq:15}
\end{align}
From relations (\ref{eq:14}) and (\ref{eq:15}), we can write
\begin{align}
a_n &= R_0\left(1 + \mu^{\textrm{rate}}\Delta - \sqrt{\frac{p}{1-p}}\nu_{\Delta}^{\textrm{rate}}\Delta\right)^n,\label{eq:16} \\
b_n &= \frac{1 + \mu^{\textrm{rate}}\Delta + \sqrt{\frac{1-p}{p}}\nu_{\Delta}^{\textrm{rate}}\Delta}{1 + \mu^{\textrm{rate}}\Delta - \sqrt{\frac{p}{1-p}}\nu_{\Delta}^{\textrm{rate}}\Delta}.\label{eq:17}
\end{align}
The coefficients in (\ref{eq:16}) and (\ref{eq:17}) represent our revised BDT model for this study.

\subsection{Simulation of Coefficients for the BDT Model}
In this study, we utilize the interest rate of the U.S. 10-year treasury bond\footnote{Related data are available on the \href{https://home.treasury.gov/}{U.S. Department of the Treasury website}.} as our primary data source. The reference starting date is set to June 16, 2023, which means that $R_0$ is set to 0.0377. Our analysis will utilize 252 days\footnote{Given that the commencement date is June 16, 2023, the data from the prior 252 days span June 15, 2022 to June 16, 2023. We have opted for a 252-day dataset because it offers a suitable sample size, ensuring that besides market fluctuations, other objective factors remain relatively stable within the year.} of historical data to compute the values of $a_n$ and $b_n$.\par
\begin{figure}[h!]
        \centering
        \includegraphics[width=0.8\textwidth]{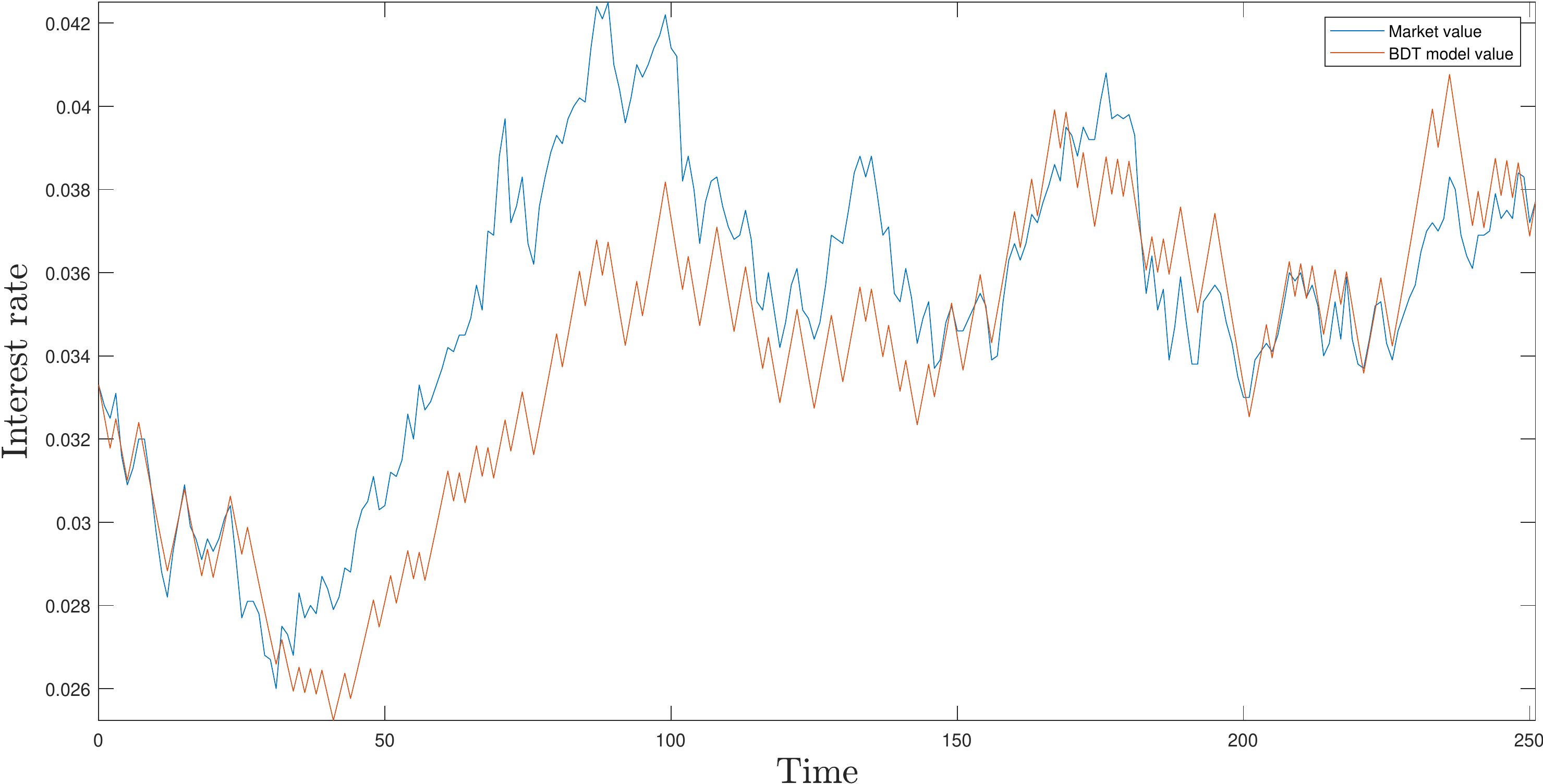}
        \caption{Comparison of market and BDT model interest rates}
        \label{BDT}
\end{figure}

Employing MATLAB for our computations, we derived the values for $c_1$ and $c_2$ as $c_1 = 1.0236$ and $c_2 = 1.0464$, respectively.
Figure \ref{BDT} contrasts the observed market value of the interest rate with the predicted values deduced from the BDT model, covering the period from June 15, 2022 to June 16, 2023.\par
Thus, for dates starting from June 16, 2023, our BDT model predicts the interest rates $R_n(\omega_1\cdots\omega_n)$ as follows:
\begin{align*}
R_n(\omega_1\cdots\omega_n) = \frac{0.0377}{1.0236^n}\cdot 1.0464^{\#H(\omega_1\cdots\omega_n)}.
\end{align*}

\section{Introduction to Zero-Coupon Bonds}
The core objective of this paper is to explore the temporal dynamics of the equity market as perceived by bond traders. To that end, this section delves into the relevant details concerning zero-coupon bonds. Subsequent subsections will first detail a pricing algorithm specific to zero-coupon bonds. Following that, we will outline a methodology for deducing the implied value of the inherent upturn probability, the instantaneous mean return, and the instantaneous volatility pertinent to the equity market.

\subsection{Zero-Coupon Bond Pricing Algorithm}
A zero-coupon bond can be viewed as a specific type of European contingent claim\footnote{We consider a zero-coupon bond to be analogous to a European contingent claim due to the intrinsic differences between the European and American trading styles. Specifically, European-style trading restricts activity until the terminal time, while American-style trading permits trading at any given moment.} (ECC). This characterization is due to its unique features: it has a predetermined maturity date, cannot be traded prior to its maturity, and guarantees a payoff of \$1 upon maturity regardless of external conditions.\par
Let us denote the price of the zero-coupon bond at time $t$, with the bond reaching maturity at time $T$, as $B(t,T)$. It is a given that $B(T,T)\equiv 1$. To price this bond, we adopt a traditional binary model, as described by \citep{shreve2004stochastic}. Using the notation established earlier,
\begin{align}
B(n\Delta, T) = \frac{1}{1+R_{n\Delta}\Delta}\left\{\tilde{p}B[(n+1)\Delta, T]^{(u)} + (1-\tilde{p})B[(n+1)\Delta, T]^{(d)}\right\},\label{eq:18}
\end{align}
where
\begin{itemize}
\item $\Delta$ represents the time interval, which is set to $1/252$.
\item $B(n\Delta,T)$ indicates the bond price at time $n\Delta$, with a maturity $T = N\Delta$.
\item $B[(n+1)\Delta, T]^{(u)}$ and $B[(n+1)\Delta, T]^{(d)}$ are potential bond prices at time $(n+1)\Delta$, with the former being greater than $B(n\Delta, T)$ and the latter being less than or equal to $B(n\Delta, T)$.
\item $R_{n\Delta}$ is the interest rate at $n\Delta$, derived using the BDT model.
\item $\tilde{p}$ is the risk-neutral upturn probability in the equity market.
\end{itemize}
For ease of representation, let us denote the bond notations $B(n\Delta, T), B[(n+1)\Delta, T]^{(u)}$, and $B[(n+1)\Delta, T]^{(d)}$ as $B_{n\Delta}$, $B_{(n+1)\Delta}^{(u)}$, and $B_{(n+1)\Delta}^{(d)}$, respectively. The instantaneous mean return in the equity market will be labeled as $\mu_{\textrm{e.m.}}$ and its variance will be labeled as $\sigma^2_{\textrm{e.m.}}$. Thus, $(\nu_{\Delta})_{\textrm{e.m.}} = \sigma_{\textrm{e.m.}}/\sqrt{\Delta}$.\par
Given the bond's ECC nature, it should be backed by an underlying risky asset $\mathbb{S}$. To ascertain the risk-neutral probabilities, one could design a risk-free portfolio:
\begin{align}
P_{n\Delta} = DS_{n\Delta} - B_{n\Delta},\label{eq:19}
\end{align}
where
\begin{itemize}
\item $P_{n\Delta}$ is the riskless portfolio value at $n\Delta$.
\item $S_{n\Delta}$ is the underlying asset's price at $n\Delta$.
\item $D$ defines the ratio of shares between the zero-coupon bond and the risky asset $\mathbb{S}$, ensuring a riskless portfolio.
\end{itemize}
Given the riskless nature of equation (\ref{eq:19}), it inherently satisfies
\begin{align*}
P_{(n+1)\Delta}^{(u)} = P_{(n+1)\Delta}^{(d)}.
\end{align*}
This leads to
\begin{align}
DS_{(n+1)\Delta}^{(u)} - B_{(n+1)\Delta}^{(u)} = DS_{(n+1)\Delta}^{(d)} - B_{(n+1)\Delta}^{(d)}.\label{eq:20}
\end{align}
Equation (\ref{eq:20}) implies the relationship
\begin{align}
D = \frac{B_{(n+1)\Delta}^{(u)} - B_{(n+1)\Delta}^{(d)}}{S_{(n+1)\Delta}^{(u)} - S_{(n+1)\Delta}^{(d)}}.\label{eq:21}
\end{align}
Given the inherent risk of asset $\mathbb{S}$ and recalling the formulation from (\ref{eq:13}), we have
\begin{align}
S_{(n+1)\Delta}^{(u)} &= S_{n\Delta}\left(1+\mu_{\textrm{e.m.}}\Delta+\sqrt{\frac{1-p}{p}}\nu_{\Delta}^{\textrm{e.m.}}\Delta\right),\label{eq:22}\\
S_{(n+1)\Delta}^{(d)} &= S_{n\Delta}\left(1+\mu_{\textrm{e.m.}}\Delta-\sqrt{\frac{p}{1-p}}\nu_{\Delta}^{\textrm{e.m.}}\Delta\right).\label{eq:23}
\end{align}
Substituting (\ref{eq:22}) and (\ref{eq:23}) into (\ref{eq:21}) yields
\begin{align}
D = \frac{B_{(n+1)\Delta}^{(u)} - B_{(n+1)\Delta}^{(d)}}{S_{n\Delta}\sigma_{\textrm{e.m.}}\sqrt{\Delta}}\sqrt{p(1-p)}.\label{eq:24}
\end{align}
From equations (\ref{eq:18}), (\ref{eq:19}), and (\ref{eq:24}), it follows that
\begin{align}
B_{n\Delta} = \frac{1}{1+R_{n\Delta}\Delta}\left[\left(p-\theta_{\textrm{e.m.}}\sqrt{p(1-p)\Delta}\right)B_{(n+1)\Delta}^{(u)} + \left(1-p+\theta_{\textrm{e.m.}}\sqrt{p(1-p)\Delta}\right)B_{(n+1)\Delta}^{(d)}\right],\label{eq:25}
\end{align}
where $\theta_{\textrm{e.m.}} = (\mu_{\textrm{e.m.}}-R_{n\Delta})/\sigma_{\textrm{e.m.}}$. From equation (\ref{eq:25}), the risk-neutral upturn probability is given by
\begin{align}
\tilde{p} = p-\theta_{\textrm{e.m.}}\sqrt{p(1-p)\Delta}. \label{eq:26}
\end{align}
With equations (\ref{eq:18}) and (\ref{eq:26}) and the condition $B(N\Delta, T) = B(T,T)\equiv 1$, we derive the complete pricing algorithm for the zero-coupon bond.

\subsection{Linking Zero-Coupon Bonds to the Equity Market}
Our objective in this subsection is to deduce the implied value of the natural upturn probability, instantaneous mean return, and instantaneous volatility for the equity market by analyzing them through the lens of a zero-coupon bond.\par
The expectation is that the zero-coupon bond's price, as calculated using the algorithm detailed in the previous subsection, aligns with the bond market's price\footnote{This price should accurately represent the zero-coupon bond's real-world value.}. To ascertain the price of the zero-coupon bond within the bond market, we refer to \citep{hull2022options} and utilize the following formula:
\begin{align}
B(t,T;\textrm{market value}) = \exp\left[-(T-t)Y(t,T)\right],\label{eq:27}
\end{align}
where
\begin{itemize}
\item $t$ represents the present time.
\item $T$ denotes the maturity date of the specific zero-coupon bond. (For robustness in our findings, we will be considering bonds with maturities spanning from two months to thirty years.)
\item $Y(t,T)$ is the yield to maturity.
\end{itemize}
The algorithm we developed in the preceding subsection provides us with a means to calculate the theoretical price of the zero-coupon bond. This can be represented as follows:
\begin{align}
B(t,T,\mu_{\textrm{e.m.}},p,\sigma_{\textrm{e.m.}};\textrm{theoretical value}).\label{eq:28}
\end{align}
Our next step entails equating equations (\ref{eq:27}) and (\ref{eq:28}). Doing this will reveal the implied values for the natural upturn probability, instantaneous mean return, and instantaneous volatility, all of which are vital metrics for understanding the equity market when it is viewed through the prism of a zero-coupon bond.

\subsection{Estimation of Parameters for the Equity Market}
To deduce the values of the parameters $\mu_{\textrm{e.m.}}$, $p$, and $\sigma_{\textrm{e.m.}}$, we will utilize the data associated with the SPY from June 15, 2022 to June 16, 2023. This duration encompasses precisely 252 business days. Our data source will be \href{https://finance.yahoo.com/quote/SPY/history?p=SPY}{Yahoo Finance}.\par
The rationale behind selecting these data is multifaceted. Bond prices often mirror the broader economic landscape. Given that the SPY can act as a representative index for the economy, it is plausible to infer that SPY data offer insight not only into the economy's state but also into the nuances of the equity market. Our preliminary estimates for $\mu_{\textrm{e.m.}}$, $p$, and $\sigma_{\textrm{e.m.}}$ are $8.0037\times 10^{-4}, 0.4821$, and $0.0126$, respectively.\par
\begin{figure}[h!]
        \centering
        \includegraphics[width=0.8\textwidth]{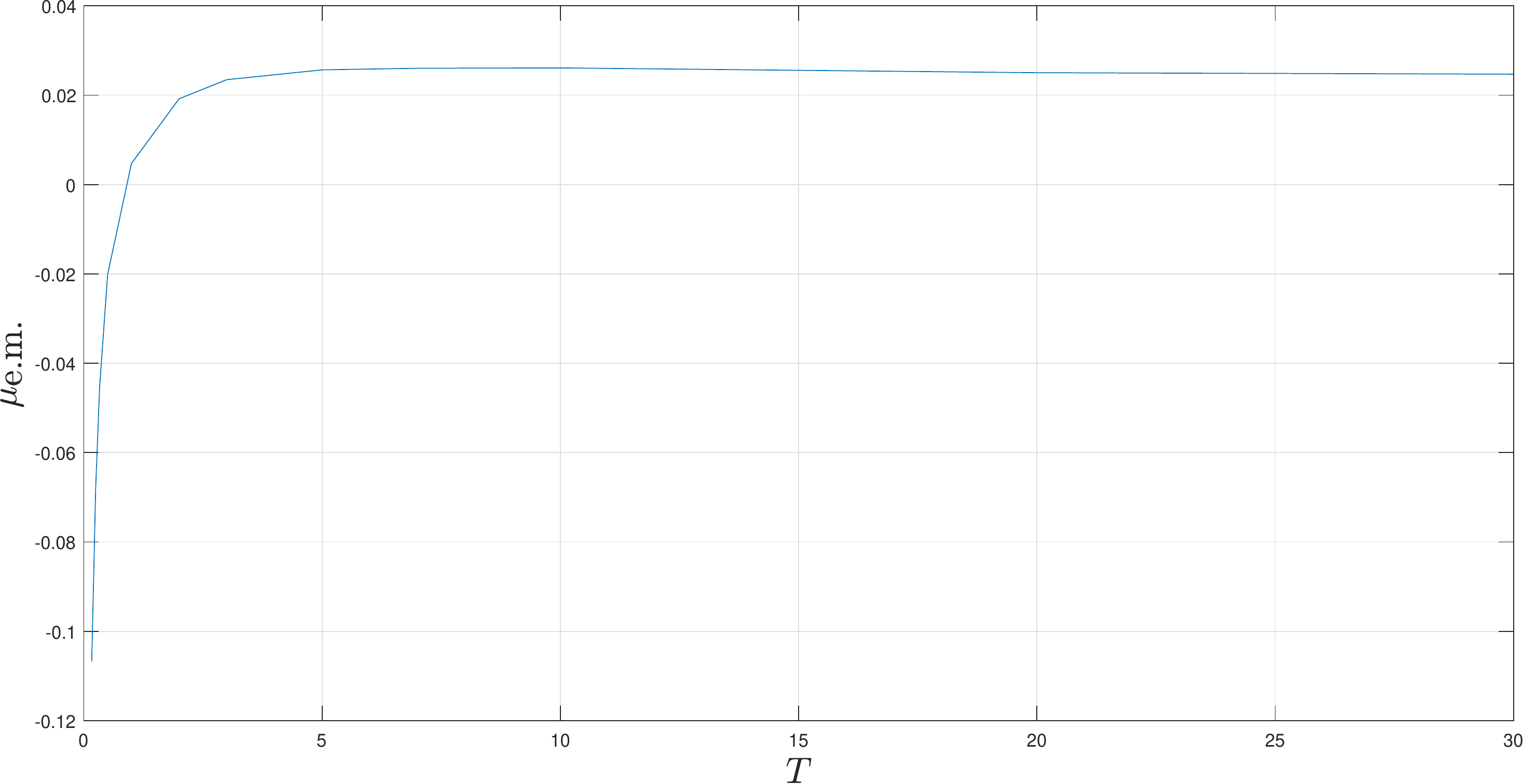}
        \caption{Temporal analysis of the implied return in the equity market}
        \label{mu}
\end{figure}
\begin{figure}[h!]
        \centering
        \includegraphics[width=0.8\textwidth]{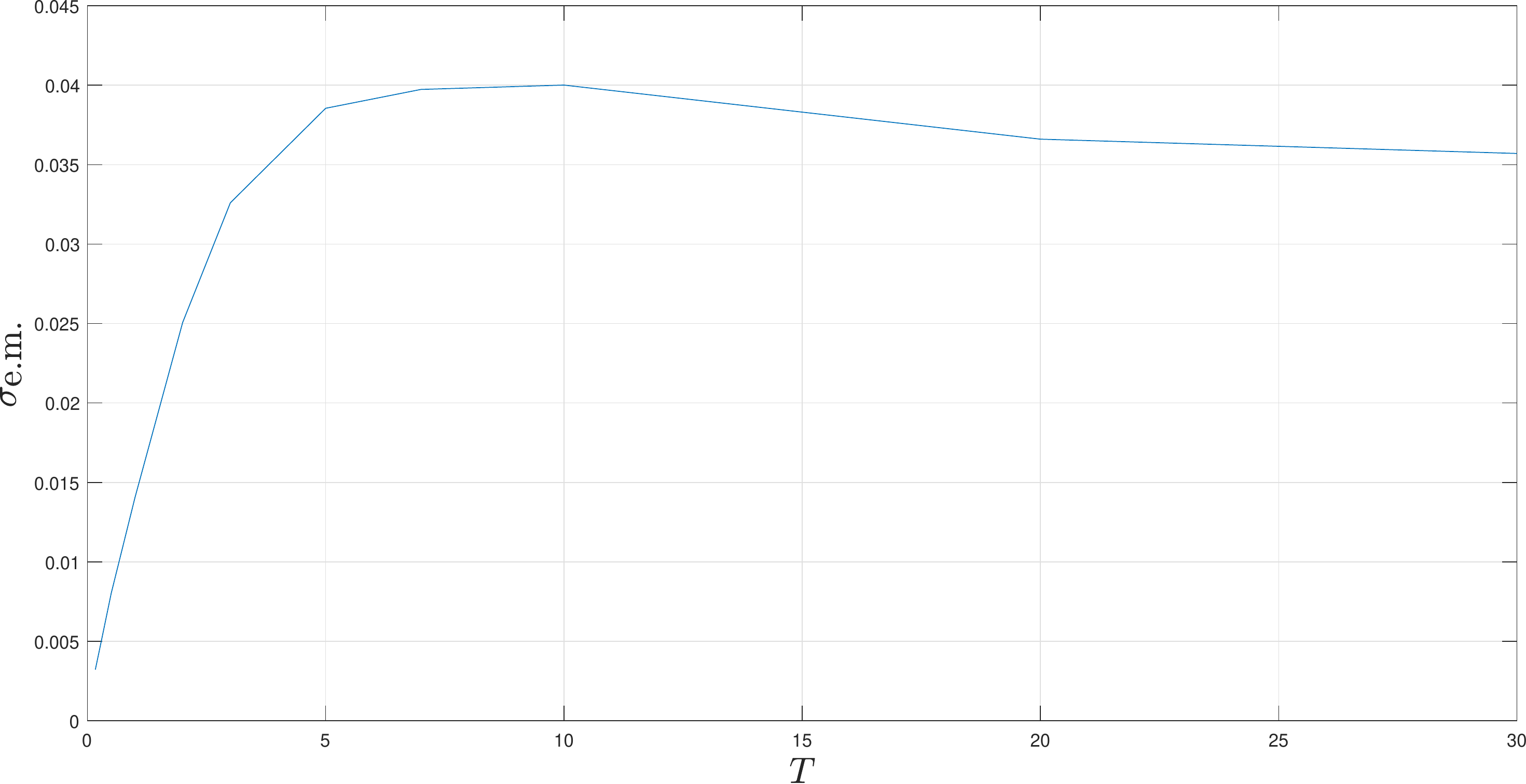}
        \caption{Temporal analysis of the implied volatility in the equity market}
        \label{sigma}
\end{figure}
\begin{figure}[h!]
        \centering
        \includegraphics[width=0.8\textwidth]{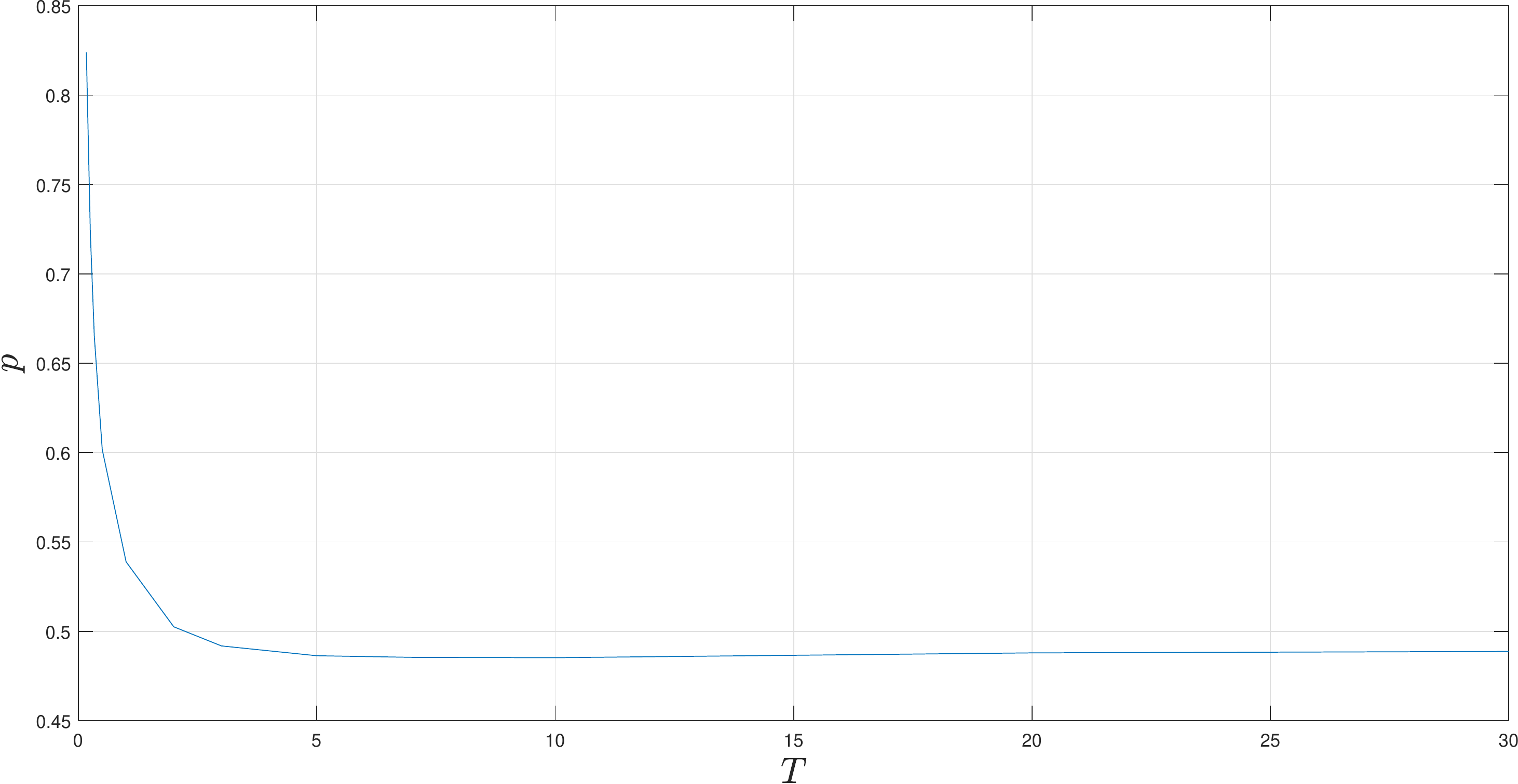}
        \caption{Temporal analysis of the implied upturn probability in the equity market}
        \label{p}
\end{figure}

As mentioned in the preceding subsection, our task is to equate (\ref{eq:27}) and (\ref{eq:28}). By doing this, we can ascertain the value of the risk-neutral upturn probability $\tilde{p}$. By leveraging equation (\ref{eq:26}) and concentrating on the parameters $p, \mu_{\textrm{e.m.}}$, and $\sigma_{\textrm{e.m.}}$, we find that any two of the parameters can be utilized to express the third. Specifically,
\begin{align}
\mu_{\textrm{e.m.}} &= \frac{\sigma_{\textrm{e.m.}}(p-\tilde{p})}{\sqrt{p(1-p)\Delta}} + R_{n\Delta},\label{eq:29}\\
\sigma_{\textrm{e.m.}} &= \frac{(\mu_{\textrm{e.m.}}-R_{n\Delta})\sqrt{p(1-p)\Delta}}{p-\tilde{p}},\label{eq:30}\\
p &= \frac{(2\tilde{p}+\Delta\theta_{\textrm{e.m.}}^2)-\sqrt{\Delta\theta_{\textrm{e.m.}}^2[\Delta\theta_{\textrm{e.m.}}^2+4\tilde{p}(1-\tilde{p})]}}{2(1+\Delta\theta_{\textrm{e.m.}}^2)}.\label{eq:31}
\end{align}
Since our start date is set to June 16, 2023, we will let $R_{n\Delta}$ be the riskless interest rate on June 16, 2023, which is $0.0377$. Now, by employing equations (\ref{eq:29}), (\ref{eq:30}), and (\ref{eq:31}), we can produce graphs reflecting the implied values for $\mu_{\textrm{e.m.}}, \sigma_{\textrm{e.m.}}$, and $p$, which are shown in Figure \ref{mu}, Figure \ref{sigma}, and Figure \ref{p}, respectively.\par
From these graphs, we can observe the following trends:
\begin{itemize}
\item In Figure \ref{mu}, the implied return rises sharply from approximately -0.11. It then starts to climb at a decelerated pace and eventually stabilizes around 0.025.
\item Figure \ref{sigma} shows that the volatility initially surges, reaching its zenith at 0.04 by the 10th year. Following this peak, there is a slight decline, and it settles near 0.035.
\item As for Figure \ref{p}, the implied upturn probability experiences a sharp dip from its starting point of roughly 0.83. This decline slows over time, with the probability ultimately converging to approximately 0.48.
\end{itemize}

\section{Conclusion}
In this study, we delved into the temporal evolution of the SPY index within the equity market through the lens of bond traders. Our focus has been on deducing the implied values of the instantaneous mean return, instantaneous volatility, and natural upturn probability over time. Our primary conclusions can be summarized as follows:
\begin{itemize}
\item The implied value of the instantaneous mean return is initially negative. However, with the passage of time, this value experiences an increase, eventually reaching a stable point.
\item As for the implied value of the instantaneous volatility, it initially has a marginal positive value. Shortly thereafter, it peaks before possibly descending to a steady state.
\item The implied upturn probability starts at a notably high value but experiences a pronounced decline as time progresses. It too attains stability over time.
\end{itemize}
\clearpage

\bibliography{mybib}

\begin{thebibliography}{7}
\providecommand{\natexlab}[1]{#1}
\providecommand{\url}[1]{\texttt{#1}}
\expandafter\ifx\csname urlstyle\endcsname\relax
  \providecommand{\doi}[1]{doi: #1}\else
  \providecommand{\doi}{doi: \begingroup \urlstyle{rm}\Url}\fi

\bibitem[Black et~al.(1990)Black, Derman, and Toy]{black1990one}
Fischer Black, Emanuel Derman, and William Toy.
\newblock A one-factor model of interest rates and its application to treasury bond options.
\newblock \emph{Financial analysts journal}, 46\penalty0 (1):\penalty0 33--39, 1990.

\bibitem[Buetow et~al.(2001)Buetow, Hanke, and Fabozzi]{buetow2001impact}
Gerald~W Buetow, Bernd Hanke, and Frank~J Fabozzi.
\newblock Impact of different interest rate models on bond value measures.
\newblock \emph{The Journal of Fixed Income}, 11\penalty0 (3):\penalty0 41--53, 2001.

\bibitem[Derman(2004)]{derman2004my}
Emanuel Derman.
\newblock \emph{My life as a Quant: reflections on physics and finance}.
\newblock John Wiley \& Sons, 2004.

\bibitem[Fabozzi(2007)]{fabozzi2007fixed}
Frank~J Fabozzi.
\newblock \emph{Fixed income analysis, 2nd edition}, volume~6.
\newblock John Wiley \& Sons, 2007.

\bibitem[Hu et~al.(2022)Hu, Lindquist, Rachev, Shirvani, and Fabozzi]{hu2022market}
Yuan Hu, W~Brent Lindquist, Svetlozar~T Rachev, Abootaleb Shirvani, and Frank~J Fabozzi.
\newblock Market complete option valuation using a jarrow-rudd pricing tree with skewness and kurtosis.
\newblock \emph{Journal of Economic Dynamics and Control}, 137:\penalty0 104345, 2022.

\bibitem[Hull(2022)]{hull2022options}
John~C Hull.
\newblock \emph{Options futures and other derivatives, 11th edition}.
\newblock Pearson, 2022.

\bibitem[Shreve(2004)]{shreve2004stochastic}
Steven Shreve.
\newblock \emph{Stochastic calculus for finance I: the binomial asset pricing model}.
\newblock Springer-Verlag, 2004.

\end{thebibliography}
\clearpage

\section*{Figure Captions}

\end{document}